# Identifying Aggregation Artery Architecture of constrained Origin-Destination flows using Manhattan L-function

Zidong Fang [a, b], Hua Shu [a, b], Ci Song [a, b], Jie Chen [a, b], Tianyu Liu [a, b], Xiaohan Liu [a, b], Tao Pei [a, b, c]∗

[a] *State Key Laboratory of Resources and Environmental Information System, Institute of Geographic Sciences and Natural Resources Research, Chinese Academy of Sciences, Beijing, China;*

[b] *University of Chinese Academy of Sciences, Beijing, China;*

[c] *Jiangsu Center for Collaborative Innovation in Geographical Information Resource Development and Application, Nanjing, China*

* Corresponding author: Dr. Tao Pei
E-Mail address: peit@lreis.ac.cn (T. Pei)
Zip code: 100101
Phone number: 01064888960

# Identifying Aggregation Artery Architecture of constrained Origin-Destination flows using Manhattan L-function

The movement of humans and goods in cities can be represented by constrained flow, which is defined as the movement of objects between origin and destination in road networks. Flow aggregation, namely origins and destinations aggregated simultaneously, is one of the most common patterns, say the aggregated origin-to-destination flows between two transport hubs may indicate the great traffic demand between two sites. Developing a clustering method for constrained flows is crucial for determining urban flow aggregation. Among existing methods about identifying flow aggregation, L-function of flows is the major one. Nevertheless, this method depends on the aggregation scale, the key parameter detected by Euclidean L-function, it doesn't adapt to road network. The extracted aggregation may be overestimated and dispersed. Therefore, we propose a clustering method based on L-function of Manhattan space, which consists of three major steps. The first is to detect aggregation scales by Manhattan L-function. The second is to determine core flows possessing highest local L-function values at different scales. The final step is to take the intersection of core flows' neighbourhoods, the extent of which depends on corresponding scale. By setting the number of core flows, we could concentrate the aggregation and thus highlight Aggregation Artery Architecture (AAA), which depicts road sections that contain the projection of key flow cluster on the road networks. Experiment using taxi flows showed that AAA could clarify residents' movement type of identified aggregated flows. Our method also helps selecting locations for distribution sites, thereby supporting accurate analysis of urban interactions.

**Introduction**

Spatial interaction, such as population migration, economic mobility, and disease spread, can be naturally abstracted into a simple origin-to-destination (OD) model called geographical flow (hereafter, "flow"). If a flow is noted as $(X_O, Y_O, X_D, Y_D)$, it can be treated as a point in a 4-D flow space, which is seen as the Cartesian product of 2-D spaces (Gao, Li, Wang, Jeong, & Soltani, 2018). Flow can be classified into two types with respect to space. One is free flow, whose origin and destination can be located anywhere within a space, e.g., the OD flow of a typhoon. The other is constrained flow, whose origin and destination can be located only in restricted locations, e.g., takeout delivery flow from restaurants to diners in road networks. Many social and economic activities in urban areas are subject to location restrictions imposed by road structures; thus, constrained flow can depict the movement of goods and people in city blocks, i.e., the travel from one place to another along roads. Analysing the patterns of constrained flow can help reveal urban dynamics.

Among the flow patterns, aggregation, viewed as the origins and destinations aggregated simultaneously, is one of the most common and significant one. Before discussing the computation of flow aggregation, we need to introduce the definition of point aggregation. In the context of point pattern, the aggregation (Kiskowski, Hancock, & Kenworthy, 2009) can be identified if the average number of points within a distance $r$ of certain point is statistically greater than that expected for a random distribution. Regarding flow, the aggregation is defined as a sphere with a radius $r$ centered within a certain flow contains more flows per unit volume than what would be expected with random flow distribution. That certain flow is termed "core flow", and the aggregation scale $r$ is the extent of the aggregation. Aggregation of human mobility flow represents intense travel demand and aggregation of item transportation flow indicates frequent freight need. Identifying and quantifying aggregation have significant functions in urban planning, such as perceiving city interactions and optimize resource allocation. Progress has been achieved for aggregated free flow by extending existing methods of detecting aggregated points; however, there is still no appropriate method for identifying constrained flow. Properly quantifying aggregated flow in constrained flow space is significant in cities (Chen et al., 2013; M. Lu, Wang, Jie, & Xiaoru, 2015; Tao & Thill, 2016).

The aggregation pattern of flows is common for heterogeneous flow data. Flow aggregation can be identified using clustering methods. Current clustering methods for flow data, whose ideas are borrowed from those of points, can be grouped into three categories (Song et al., 2019). The first is the hierarchical clustering method proposed for cluster identification and visualization (Adrienko & Adrienko, 2011; Doantam, Ling, Yeh, & Hanrahan, 2005; Lee, Han, & Whang, 2007; Wood, Dykes, & Slingsby, 2010). This approach can classify flows into clusters with multiple scales according to the parameters set for cluster numbers (X. Zhu & Guo, 2014). However, selecting the optimal parameter is problematic, and it is sensitive to noise: it cannot eliminate noises and even an outlier flow would be included in a certain cluster.

To address this noise problem, the density-based method has been proposed to treat flow data (Nanni & Pedreschi, 2006; Pei et al., 2015; B. Zhu, Huang, Guibas, & Zhang, 2013). This method is typically developed by altering the distance, density, and reachability using the density-based spatial clustering of applications with noise (DBSCAN) algorithm. DBSCAN can detect arbitrarily shaped clusters and reject outlier flows; however, determining the parameters $MinPts$ and $\varepsilon$ is complex. It's hard to determine the key flows in the cluster, which is thereby impossible to recognize the key characteristics of this aggregation.

In light of the above problems, the spatial statistic-based method was developed; it extended existing spatial statistics, such as Moran's I (Liu, Tong, & Liu, 2015), Getis-Ord G statistics (Berglund & Karlström, 1999), and Ripley's K-function (Tao & Thill, 2016). The main purpose with this method is to create a new significant test for spatial homogeneity. If the null hypothesis is rejected, a global or local flow clustering anomaly is detected (Y. Lu & Thill, 2003). Among existing spatial statistic-based methods, L-function, which is developed from K-function, is the top choice for identifying and quantifying aggregation: it can reveal the centre and scale of aggregation without manually setting parameters while its local version can extract clusters based on corresponding scales (Haase, 1995; Sterner, Ribic, & Schatz, 1986). To discern the aggregation scales and extract corresponding flow clusters, (Shu et al., 2020) proposed a Euclidean distance-based L-function for geographical flows to estimate the aggregation extent and identify the dominant clusters efficiently, which cannot be solved by previous methods.

Nevertheless, the existing flow L-function is not appropriate for the flow study in urban situation, which is constrained by road networks. First, the aggregation scale detected with Euclidean distance is inconsistent with the city environment. Although scholars tried to extend K-function or L-function with true-to-life distance measurements for point pattern analysis (Lamb, Downs, & Lee, 2016; Okabe & Yamada, 2001) and for flow pattern analysis (Kan, Kwan, & Tang, 2021), lack of theoretical basis in flow space still led to untrue aggregation scales. Since cities are usually divided into blocks with such functional areas as commercial districts, medical centres, and city halls. Thus, when measuring the distance between two places in cities, it is often necessary to consider how many blocks they are apart (i.e., Manhattan distance). Therefore, Manhattan distance may operate better since it is not only not only closer to the real road network distance, but also more concise to derive theoretical properties of flow space, thus providing more truthful information.

The second flaw of Euclidean L-function is that the clustering results are too scattered to reflect crucial characteristics of aggregation in the road network. For example, by detecting the aggregation of taxi OD flow and analysing urban function of its ends, we could deduce the movement type of the passengers. However, due to poor concentration of aggregation identified by existing L-function, it is hard to determine its movement type. Therefore, Aggregation Artery Architecture (AAA), depicting core distribution extent of aggregation in road networks, is proposed in our study to reveal key information of aggregations. However, it is inappropriate to obtain AAA by overlay analysis of flow aggregation and road networks, namely filtering out the aggregated flows that located on road networks, because not every flow which is close to the aggregation centre along the road networks should be included in AAA. Finding out the key flows of aggregation from the perspective of spatial statistics would be the right way and the intersection of important flow sets can reflect the most key part of the aggregation.

In the present study, we define "Manhattan flow space" as a flow space based on Manhattan distance, which adapts well to flows in road networks. After deriving benchmark of K-function by defining complete spatial randomness of flows in Manhattan space, we propose the corresponding L-function of flows to identify the aggregation centre and its extent easily and accurately. By taking the intersection of several key clusters decided by L-function, we could determine AAA, the flows on which locate along or near the same roads. Such aggregation characteristics of constrained flows as centre, scale, and AAA, they all play significant roles in understanding aggregation phenomena in cities, thus supporting more reasonable urban planning.

*Concepts*

*Basic measurements in Manhattan flow space*

**Definition 1** Maximum Manhattan distance: "Manhattan distance" is also termed "city block distance" and "taxicab distance" (Krause, 1986). It can roughly depict the distance along a road network. In the 2-D plane, the Manhattan distance between point $(x_i, y_i)$ and point $(x_j, y_j)$ is denoted as $d_{ij}$:

$$d_{ij} = |x_i - x_j| + |y_i - y_j| \qquad (1)$$

A geographical flow, $F_i$, can be represented by the ordered combination of rectangular coordinates of point $O(x_i^O, y_i^O)$ and point $D(x_i^D, y_i^D)$, i.e., $F_i\left((x_i^O, y_i^O), (x_i^D, y_i^D)\right)$. The distance between flow $F_i$ and flow $F_j$ can be denoted as $D_{ij}$. In previous studies, the distance between flows, namely $D_{ij}$, has usually been designed to be the linear combination or extreme value of $d_{ij}^O$ and $d_{ij}^D$, where $d_{ij}^O$ represents the distance between $O$ points and $d_{ij}^D$ signifies the distance between $D$ points (Shu et al., 2020; Tao & Thill, 2016). In different application scenarios, it is possible to use different means to measure the proximity of flows. The additive distance and maximum distance are two frequently used measurements. The former equals the weighted sum of $d_{ij}^O$ and $d_{ij}^D$; the latter is the larger value of $d_{ij}^O$ and $d_{ij}^D$. Here, we denote the maximum distance as $D_{ij}^{(max)}$. In this study, the maximum distance can reflect the extent of aggregated flows appropriately: it ensures that both ends of a flow are limited in a certain distance to the corresponding ends of another flow. Therefore, the following measurements about proximity are founded on maximum Manhattan distance (Figure 1a):

$$D_{ij}^{(max)} = \max(d_{ij}^O, d_{ij}^D) \qquad (2)$$

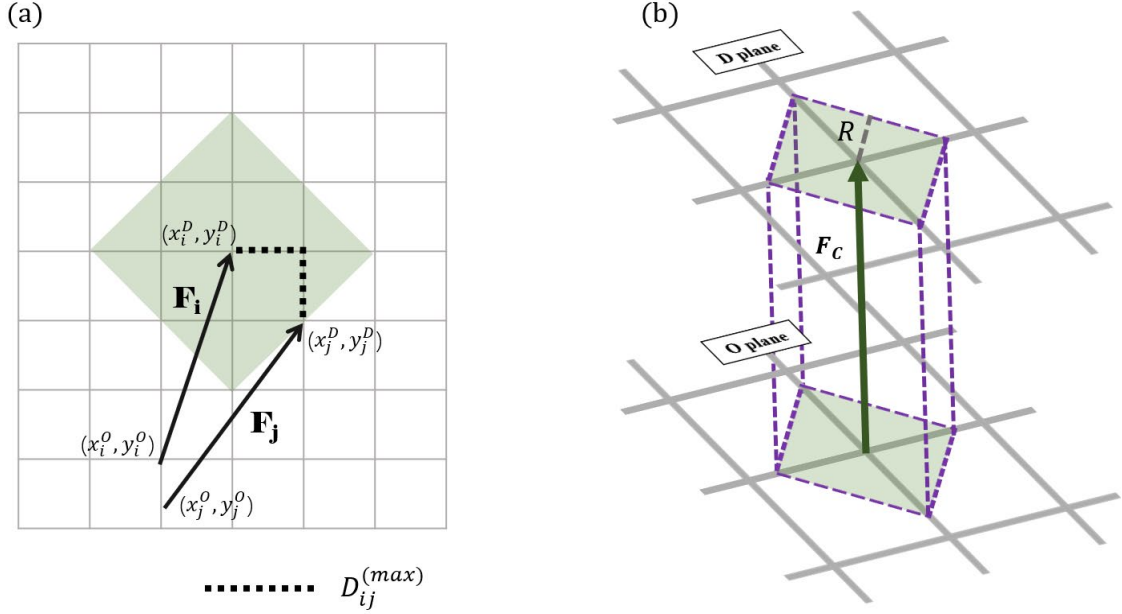

Figure 1 Flows in Manhattan flow space with maximum distance: (a) maximum distance between flows; (b) projection of a flow sphere in 4-D flow space.

**Definition 2** Flow volume: In a flow space, suppose the area of the O plane to be $S_O$ and the area of the D plane to be $S_D$; the volume of this flow space should be $V = S_O S_D$. As an example, we apply this sphere in Manhattan flow space. As shown in Figure 1b, given a sphere $\Omega$ with radius $R$ in Manhattan flow space based on centre flow $F_C$, the flows in $\Omega$ are those within $R$ maximum Manhattan distance of $F_C$, i.e., $\Omega = \{F | D_{F,F_C}^{(max)} \leq R\}$. The volume of sphere $\Omega$ can be derived as

$$V_\Omega^{(max)} = SO \cdot SD = 4R^4 \qquad (3)$$

**Definition 3** Flow process intensity: Flow process intensity is the expected number of flows per unit volume. Although (Kan et al., 2021) proposed a method which uses the length of the road network in the experimental area to represent support domain, it is unreasonable because flow space is the Cartesian product of two 2-D planes (Gao et al., 2018), which will seriously affect the judgment of flow aggregation and corresponding scales.

In the real world, although it is difficult to identify precisely the boundary of the support domain with flow process; the intensity parameter can be estimated using the second-order properties of flows (Pei et al., 2015; Pei, Zhu, Zhou, Li, & Qin, 2007); that is calculated with $d_{i,1}$, the first-order nearest-neighbour distance to each flow:

$$\hat{\lambda}_F = \frac{1}{4E(d_{i,1}^4)} = \frac{n}{4\sum_{i=1}^n d_{i,1}^4} \tag{4}$$

Here, the first-order nearest-neighbour distance equals the minimum distance from one flow to another in this flow dataset; $n$ is the number of flows in the research domain.

*Aggregation Artery Architecture*

AAA is the road sections that contain the projection of key flow cluster on the road networks; it could reflect the core information about the aggregation. For example, in Figure 2, the ground-truth is that the aggregation is from functional area A to B. But the aggregation identified by Euclidean L-function may contains some noises (red flows, from functional area C to D), they mislead the understanding about movement type of this mobility aggregation. The aggregated flows on AAA are from functional area A to B, we can thereby counteract the influence of red on the judgment of main aggregation.

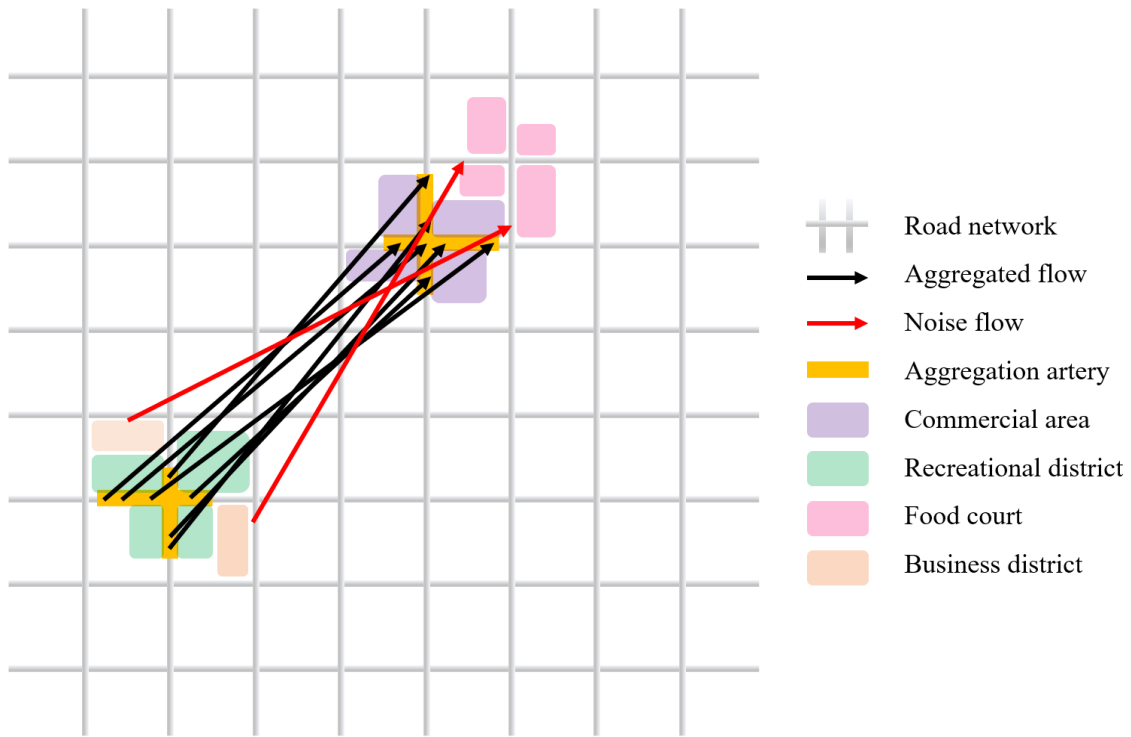

Figure 2 AAA of taxi OD flows in a road network. The red flows have a different movement type because the urban function of their origins and destinations differ from those of the dominant flows on AAA (colored black); the dominant flow in this cluster is from the recreational district to the commercial area; noise flows are from the business district to the food court.

### *L-function based on Manhattan distance*

In this section, we derive the flow K-function and complete spatial randomness (CSR) in Manhattan flow space. The L-function, depicting the deviation from CSR distribution, and its local version are also illustrated for detecting scales and clustering. The technique of using parameter $T$ (number of selected core flows), which we designed for clustering flows with the same travel range, is introduced at the end of this section.

*K-function*

As with Ripley's K-function in point-pattern analysis, the K-function for flows is defined as the expected number of additional flows within a maximum Manhattan distance, $r$, of an arbitrary flow normalized by flow process intensity. The derived formula for $K(r)$ is

$$K(r) = \lambda_F^{-1} \frac{\sum_i \sum_j \sigma_{ij}^F(r)}{n}, (i, j = 1, 2, \ldots, n; i \neq j) \qquad (5)$$

$$\sigma_{ij}^F(r) = \begin{cases} 1, d_{F_i,F_j}^{(max)} \leq r \\ 0, d_{F_i,F_j}^{(max)} > r \end{cases} \qquad (6)$$

where $r$ is a designated maximum Manhattan distance, $n$ is the number of flows in the research area, and $\lambda_F$ is the intensity of the flow process.

*L-function and local L-function*

In point-pattern analysis, CSR describes a point process whereby point events occur within a given study area in a completely random distribution. CSR is often applied as the standard or benchmark against which datasets are tested. In the present study, inferences about CSR flows assist in deriving the L-function from the K-function. CSR flows are defined as a homogeneous spatial Poisson process. Given a subset $S$ of flow space $\Psi$, the number of flows in $S$ obeys the Poisson distribution as $\lambda_F V_S$, where $\lambda_F$ is the intensity of this flow process and $V_S$ is the volume of flow space $S$ (Shu et al., 2020).

As defined in section 3.1, the expected number of additional flows within a maximum Manhattan distance $r$ is $\lambda_F K(r)$. Combining that with the property of CSR flows, we can conclude that $\lambda_F K(r) = \lambda_F 4r^4$. Thus, the expectation for the K-function with a homogeneous Poisson flow process is

$$E\big(K(r)\big) = \frac{\lambda_F 4r^4}{\lambda_F} = 4r^4 \qquad (7)$$

By normalizing the K-function with the expectation for the K-function, we can obtain the L-function: the expectation is 0 at any scale (distance) for CSR flows. In this way, we can analyse the deviation from flow CSR:

$$L(r) = \sqrt[4]{\frac{K(r)}{4}} - r = \sqrt[4]{\frac{\sum_i \sum_j \sigma_{ij}^F(r)}{4\lambda_F n}} - r, (i,j = 1,2,\ldots,n; i \neq j) \qquad (8)$$

The local L-function for each flow is defined as

$$L_i(r) = \sqrt[4]{\frac{\sum_j \sigma_{ij}^F(r)}{4\lambda_F n}} - r, (i,j = 1,2,\ldots,n; i \neq j) \qquad (9)$$

*Extracting AAA*

As explained before, AAA is the road sections that contain the projection of key flow cluster on the road networks. Extracting AAA consists of two parts, the first one is about identifying aggregations and the second one is concentrating them as a key cluster to highlight AAA.

First, with the aggregation scale, $\hat{R}$, provided by the L-function, we can calculate the value of the local L-function, $L_i(\hat{R})$, for each flow in the dataset. Then, we select the flows with the highest local L-function values and treat them as core flows at this scale. Here, we denote the number of core flows as $T$. Then, we identify core flows and their $\hat{R} - neighbors$, i.e., flows within $\hat{R}$ to those core flows. Finally, we regard the intersection of clusters dominated by core flows as the key cluster and project it on the road networks for subsequent sketching. In a word, finding a concentrated flow cluster in road network is the key to figure out AAA.

Different from conventional clustering algorithm (e.g., DBSCAN), which search and combine the neighbourhood of decisive flows, our method includes only the intersection of $\hat{R} - neighbors$ into the clusters. A simple example appears in Figure 3. Suppose aggregation scale is $\hat{R}$ and we chose $T = 2$. The flows with the black stars are the core flows; the green spaces are the $\hat{R} - neighborhoods$ of those two core flows. The dark green space is the intersection of those two $\hat{R} - neighborhoods$. In this case, only the orange flows at the intersection are included in the key cluster; the blue flows are regarded as noise. The O points in this example are close to one another: the maximum distances between the flows are the distances between the D points. Thus, we need examine only the D plane.

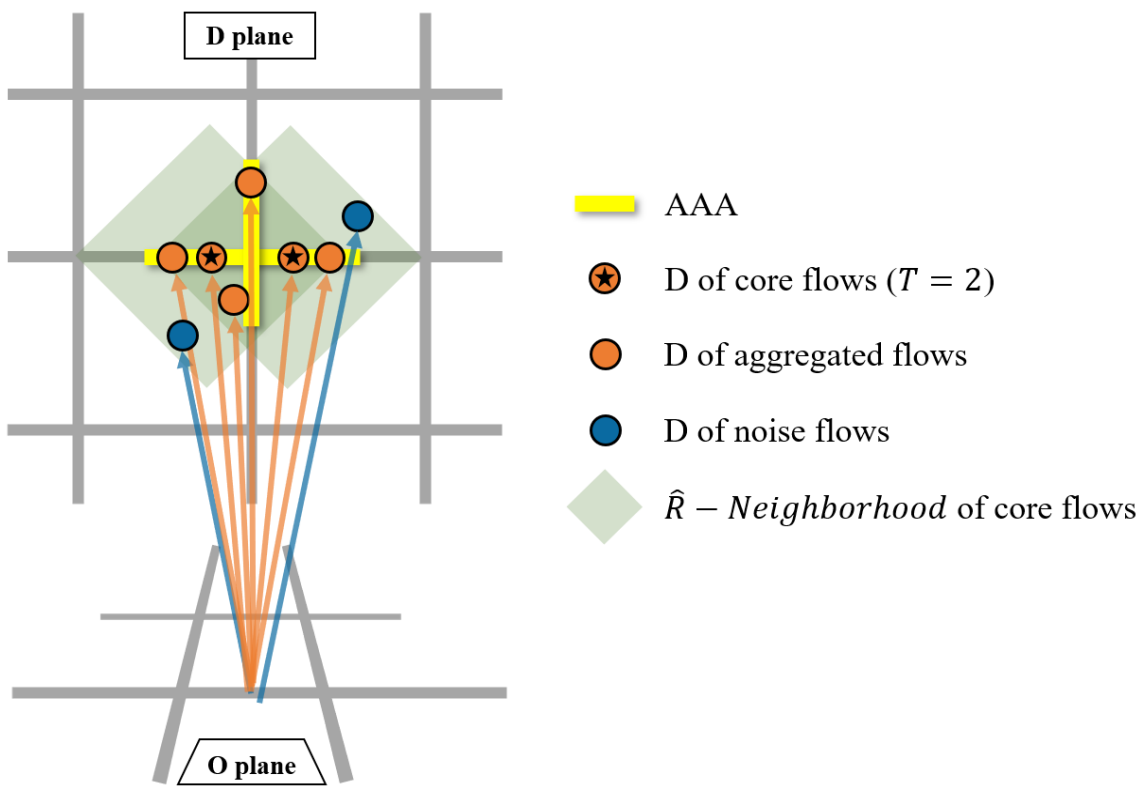

Figure 3 Illustration of extracting AAA. In this case, $T = 2$, i.e., it picks top two core flows; the intersection of their *R-neighborhoods* will be the clustering range. Thus, the orange flows are in the key cluster; the blue flows are rejected. Highlight AAA by sketching the boundary of projection of key cluster on the network.

To a certain extent, by setting the value of $T$, we can guarantee that the flows contained in the key cluster are within Manhattan distance of $\hat{R}$ to the centre flow (Figure 3). With the practical example of taxi OD flows, this key technique ensures that the driving distance from each flow to the centre flow in the cluster does not exceed $\hat{R}$; that helps overcome the problem of clustering unnecessary flows in urban environment.

Here, we conduct a comparison experiment about clustering results with different $T$ values (Figure 4). Noise flows whose ends cannot access the corresponding ends of the centre flow within Manhattan distance of the aggregation scale are eliminated gradually with increasing $T$ values. The number of aggregated flows in the experimental dataset is 300. When $T = 21$, this method attains optimal performance when the number of detected aggregated flows is 301.

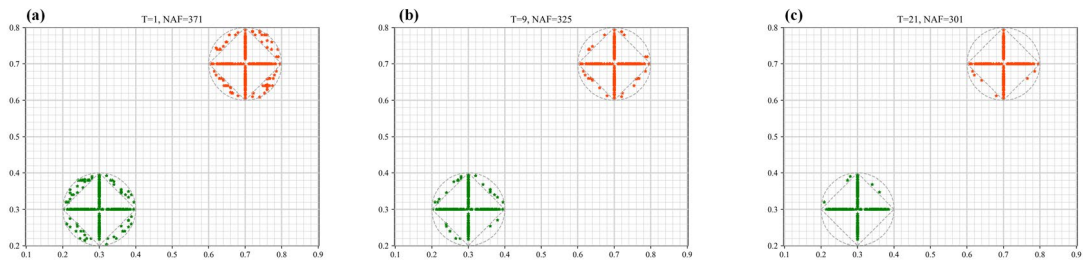

Figure 4 Extraction results with different $T$ values. The green points and orange points represent the O points and D points, respectively. $T = 1$ in (a); $T = 9$ in (b); $T = 21$ in (c); NAF, number of aggregated flows in the result.

*Simulation experiment*

The organization of urban street networks is too complex for description; however, the basic forms can be examined as grid patterns because many cities are divided into blocks. Grid patterns (also called rectangular or block patterns) are where the roads run at right angles to one another, forming a grid. This kind of network appears in such places as Beijing, Xi'an, and New York.

To test the concentrating ability of the Manhattan L-function, we constructed eight types of flow datasets (Figure 5); they approximately simulate the common situation with urban roads. All the flows in the datasets are constrained by roads: the O points and D points of the flows are located on the designed networks. Each flow dataset comprises 300 aggregated flows, 100 noise flows, and 300 background flows. The centre flow of the aggregated flows is $((0.3,0.3),(0.7,0.7))$, and the aggregation scale is 0.1. The aggregated flows are those that accumulate on the main roads (thick grey lines) in the designed patterns; the noise flows appear on secondary roads (fine grey lines) within Euclidean distance of $R$ to the aggregated center; and the background flows are randomly arranged in roads.

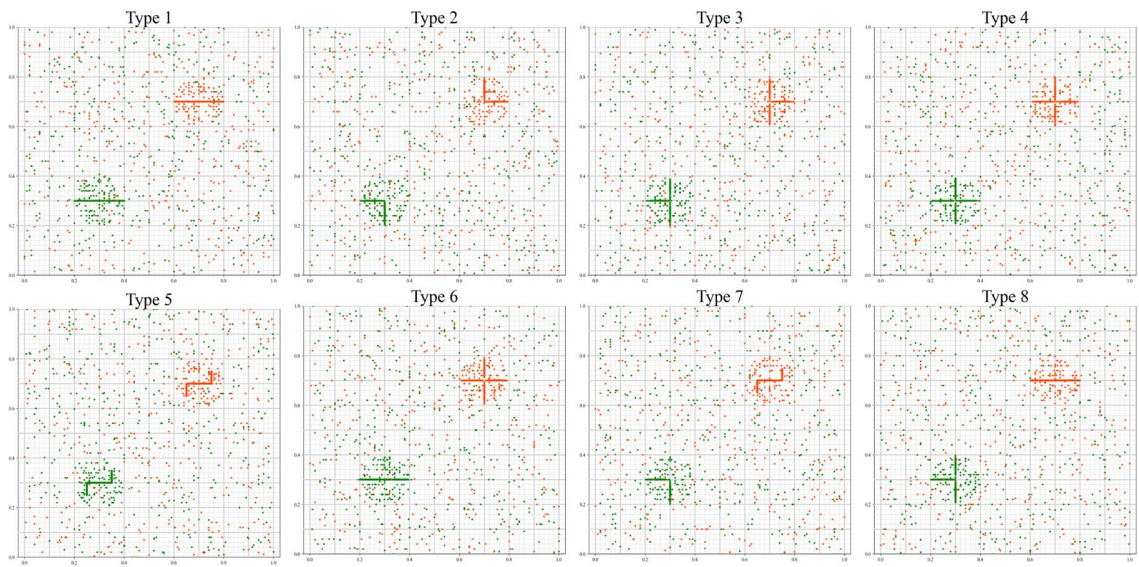

Figure 5 Simulated flow data; O points appear in green and D points are in orange.

With the aggregation scale $R$, the objective cluster should contain aggregated flows on the main roads (thick grey lines). We utilized standard evaluation metrics—precision, recall, and F1 score (i.e., the weighted average of precision and recall)—to compare the clustering ability among the L-function with Manhattan distance, L-function with Euclidean distance, and Manhattan flow DBSCAN ($\varepsilon = R, MinPts = 10$).

The results with the three different clustering methods for the type-4 flow dataset appear in Figure 6. It is evident that the clusters extracted by Manhattan flow DBSCAN and L-function with Euclidean distance (which contain some noise flows far from the cluster centre) are not concentrated compared with the result cluster by L-function with Manhattan distance, which shows that the result flow cluster of our method are more concentrated.

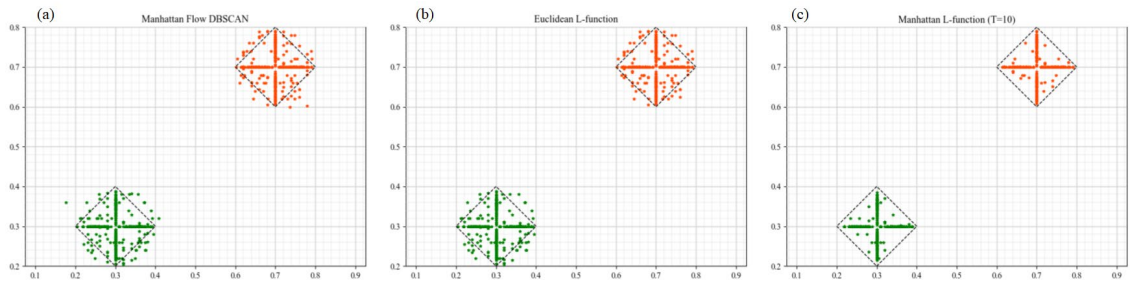

Figure 6 Results with different clustering methods: (a) Manhattan flow DBSCAN; (b) Euclidean L-function; (c) Manhattan L-function. The O points appear in green and the D points in orange. The OD area within the dotted lines is the $\hat{R} - neighbors$ of the centre flow.

For these eight types of flows, we conducted tests with 100 random datasets in the corresponding patterns. The average results for the detected aggregation scale and the standard evaluation metrics appear in Table 1. It is evident that the L-function with Manhattan distance yielded an average F1 score greater than that with the other two clustering methods. Thus, the Manhattan L-function was better at clustering the flows that are concentrated on the main road and thus sharing same flow properties, which is beneficial for extracting AAA.

Table 1 Results of the simulation experiment

| | Manhattan Flow L-function | | | | Euclidean Flow L-function | | | | Manhattan Flow DBSCAN | | |
|---|---|---|---|---|---|---|---|---|---|---|---|
| OD Type | Scale | Precision | Recall | F1-score | Scale | Precision | Recall | F1-score | Precision | Recall | F1-score |
| 1 | 0.12 | 0.86 | 0.97 | 0.92 | 0.11 | 0.75 | 1.00 | 0.86 | 0.75 | 1.00 | 0.85 |
| 2 | 0.12 | 0.90 | 0.92 | 0.91 | 0.10 | 0.76 | 1.00 | 0.87 | 0.75 | 1.00 | 0.85 |
| 3 | 0.12 | 0.91 | 0.91 | 0.91 | 0.10 | 0.76 | 1.00 | 0.86 | 0.74 | 1.00 | 0.85 |
| 4 | 0.12 | 0.91 | 0.91 | 0.91 | 0.10 | 0.75 | 1.00 | 0.86 | 0.74 | 1.00 | 0.85 |
| 5 | 0.12 | 0.87 | 0.94 | 0.90 | 0.09 | 0.77 | 1.00 | 0.87 | 0.74 | 1.00 | 0.85 |
| 6 | 0.12 | 0.89 | 0.94 | 0.91 | 0.10 | 0.75 | 1.00 | 0.86 | 0.75 | 1.00 | 0.85 |
| 7 | 0.12 | 0.89 | 0.93 | 0.90 | 0.10 | 0.76 | 1.00 | 0.86 | 0.74 | 1.00 | 0.85 |
| 8 | 0.12 | 0.89 | 0.95 | 0.92 | 0.10 | 0.75 | 1.00 | 0.86 | 0.74 | 1.00 | 0.85 |
| Average | 0.12 | 0.89 | 0.93 | 0.91 | 0.10 | 0.76 | 1.00 | 0.86 | 0.74 | 1.00 | 0.85 |

*Experiment with real data*

For the research area, we selected the Shunyi district, which is a subcentre of Beijing and near the inside of the eastern Sixth Ring Road. The road network in this area is roughly in grid pattern, it can well demonstrate the effectiveness and robustness of our method. For this area, we chose the taxi OD flows for October 20–24, 2014. We examined flows that exceeded 2 km and we identified 1081 taxi OD flows in the district for our experiment (Figure 7).

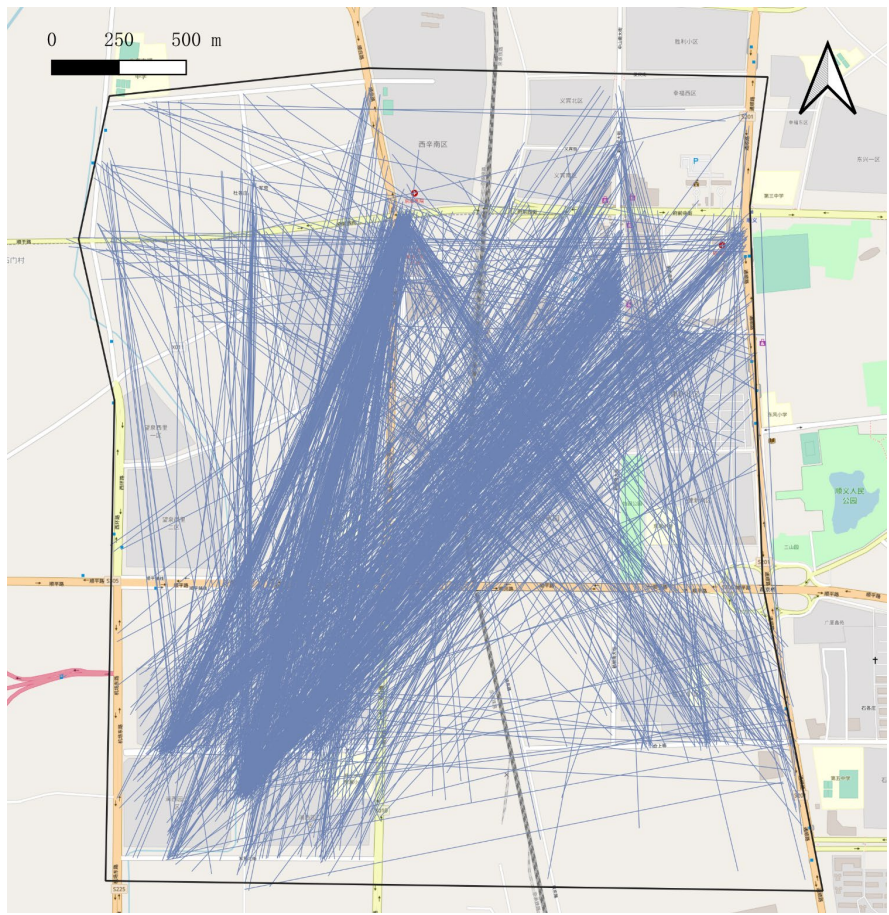

Figure 7 Research area (Shunyi district) and the selected taxi OD flows.

In the following analysis, we examine the L-function of flows in Manhattan space (MLF) and L-function of flows in Euclidean space (ELF). Combining $\arg[L_{max}(r)]$ ($r$ value that makes the L-function attain the maximum) and $\arg[L'_{min}(r)]$ ($r$ value that makes the derivative of the L-function reach the minimum), the maximal aggregation scale was 110 m detected by ELF (Figure 8b) and 150 m by MLF (Figure 9b). After the maximal aggregation scale, the remaining minima of $L'(r)$ represent aggregations in different scales. However, aggregation extracted using large scales are typically combinations of individual aggregations extracted by several small scales. Here, we focus on the main individual aggregations, i.e., maximal and secondary aggregation. For better analysis, we compared the aggregations extracted using Manhattan flow DBSCAN ($\varepsilon = 250, MinPts = 20$), ELF, and MLF (Figure 10). We found that the aggregated flows with MLF were more concentrated and shared a similar range in the road network, thereby providing a clear picture of the movement type.

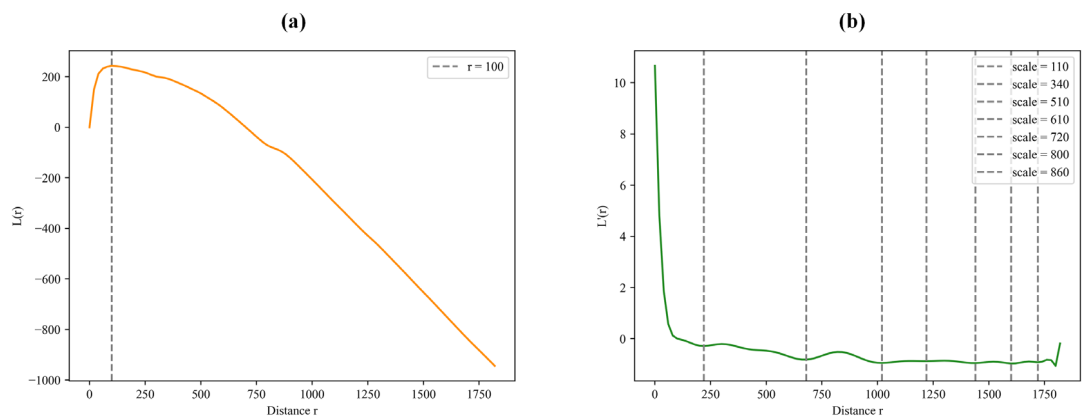

Figure 8 Results of the study case with Euclidean L-function: (a) L-function; (b) derivative of L-function.

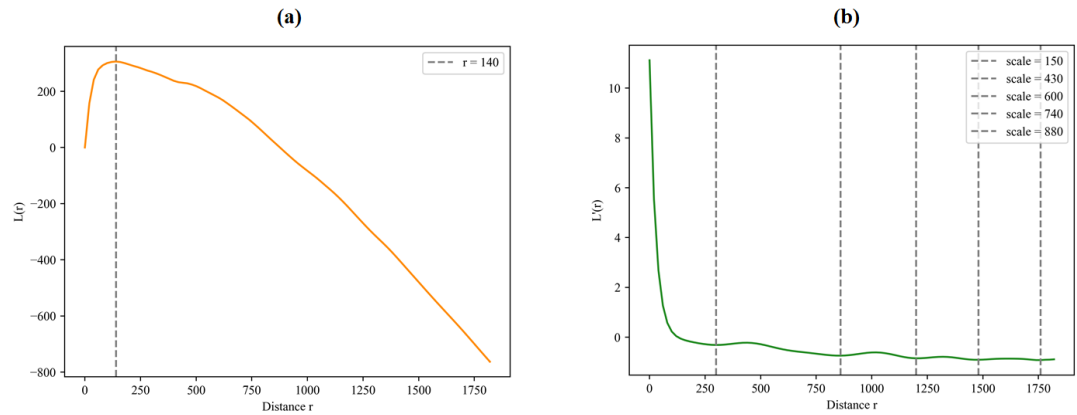

Figure 9 Results of the study case with Manhattan L-function: (a) L-function; (b) derivative of L-function

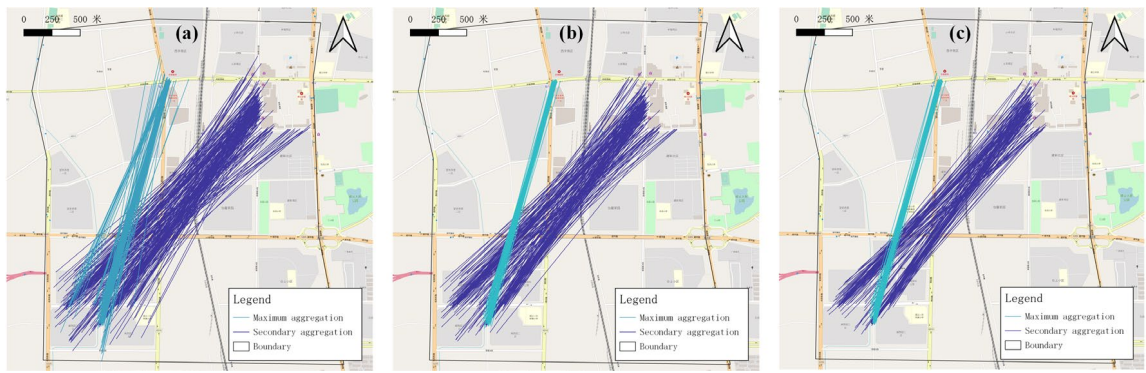

Figure 10 Clustering results of the maximal aggregation and secondary aggregation: (a) result with Manhattan flow DBSCAN; (b) result with ELF; (c) result with MLF ($T = 3$)

There was little difference between the maximal aggregation extracted using the three methods. We illustrate here the effectiveness of our method with secondary aggregation. The aggregation range with Manhattan flow DBSCAN was broad compared with the other two methods: the maximum Manhattan distances between origins and destinations were too large to determine whether the origins or destinations were in the same functioning district. Thus, we focus here on the difference between secondary aggregation extracted using ELF and MLF. Figure 12 shows the difference between the two clusters: the flows were clustered with ELF but rejected with MLF.

After sketching the AAA of aggregation using MLF, we could observe a clear movement type (Figure 11). The origins of the AAA were all located on Xin Shun South Street, which is a commercial thoroughfare composed of several shopping centres. However, the origins of the cluster extracted with ELF (Figure 12)—some of which were located near banks, bookstores, or hospitals—were not concentrated in the same functioning block. We obtained the same result with destinations for D points of AAA: they were all located in a residential district (Lanxi Garden); the flow cluster extracted using ELF revealed some destinations for clinics, a technology museum, and office buildings. The reason for the cluster result with ELF being dispersed compared with MLF was that some flows that were close to the aggregation centre using Euclidean distance would be included in the ELF result; examples here are the flow from the bank to the clinic and flows from the commercial area to business buildings. The MLF would reject those flows because they are somewhat distant from the center in Manhattan space. Thus, with the flow aggregation MLF, it is easy to extract AAA, revealing a definite moving purpose of those passengers (i.e., from the commercial area to the residential district); that did not become evident with the equivocal ELF result, which indicated imprecise, un-concentrated clustering.

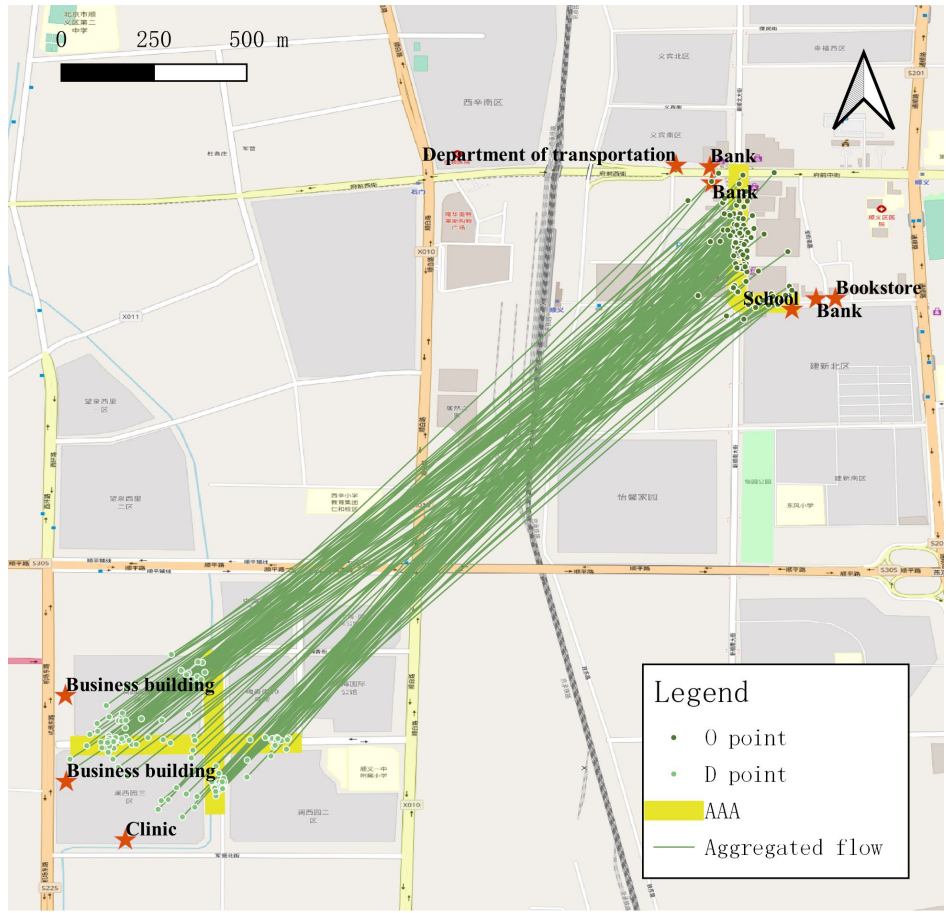

Figure 11 Secondary aggregation obtained using MLF and corresponding AAA; the flows in this cluster are mostly from the commercial area to the residential district.

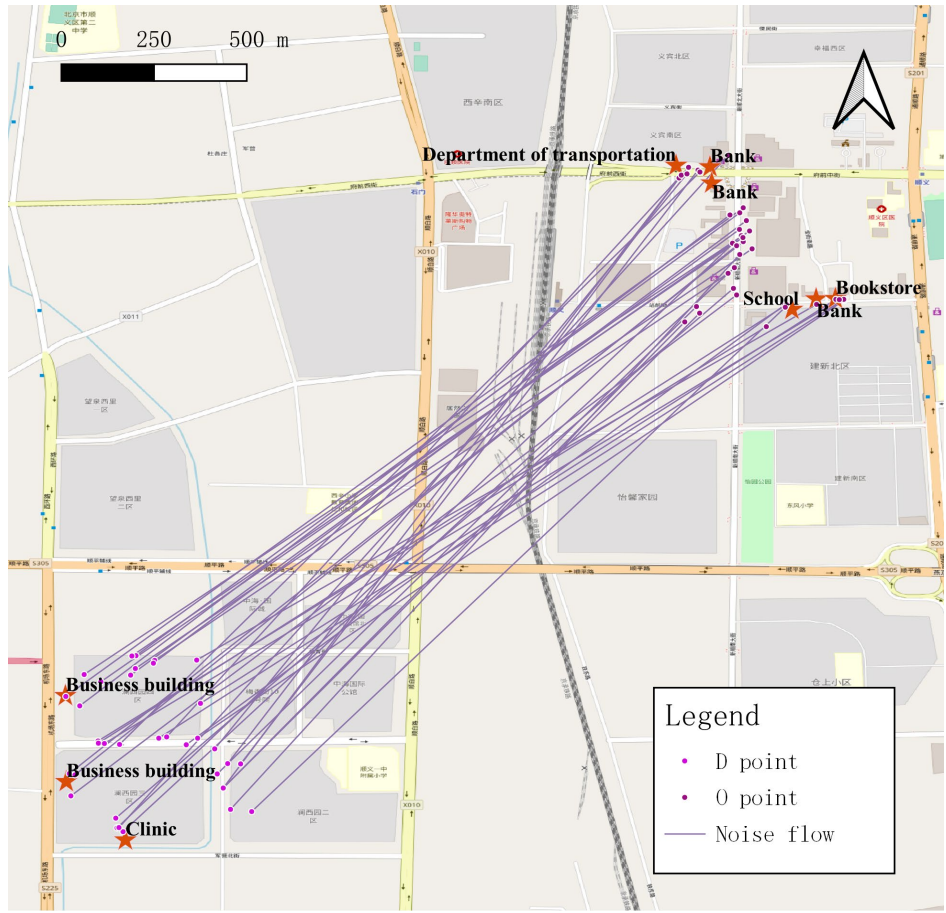

Figure 12 Difference between the secondary aggregation obtained using ELF and MLF; the noise flows clustered with ELF were rejected with MLF. Some of these flows were from banks, schools, or bookstores and ended near clinics or business buildings, thereby confusing the main direction with this flow cluster.

## *Conclusions*

To identify the aggregation pattern of constrained flows and analyze its characteristics, we constructed a Manhattan flow space; we proposed a flow clustering method by improving the existing flow L-function with Manhattan distance. This method can detect more concentrated aggregations over different scales using the L-function and extract the Aggregation Artery Architecture, revealing most crucial information. In that way, it assists in analysing moving patterns in cities: it offers practical benefits to urban research.

The L-function plays a key role in our method: it can quantify a range of aggregation scales by measuring the level of deviation from flow CSR in Manhattan space; its local version can also cluster corresponding aggregated flows. Thus, with the aid of the L-function, our method can automatically identify the location and extent of aggregations in flow datasets without setting parameters involving multiple trials. In addition, since more concentrated and structured aggregation cannot be derived by existing Euclidean L-function, extracting AAA is a typical method in Manhattan flow space.

The results with our method conformed well to the actual urban situation. The aggregation scale detected using Manhattan L-function was larger and more practical than that with Euclidean L-function. That is because when estimating the distance between two urban locations, it may not be possible to measure the straight-line distance between the two points directly: it is necessary to consider how many blocks apart they are. Accordingly, for flows in road networks, the scale provided using Manhattan L-function is more realistic and informative. Further, flows located on AAA, which is extracted by the aid of more concentrated flow aggregation (optimized with parameter $T$), share the same travel range in blocks. That ensures there is a high probability of the origins or destinations of the cluster being in the same intersection or road within the city, which enhances the capturing of city dynamics and supports appropriate urban planning.

Our improved L-function has some shortcomings. The method we propose has requirements concerning the structural characteristics of the road network. It performs better with grid-like networks or cities that are divided into blocks. The $T$ value (discussed in detail in section 3.3) plays an essential role in extracting clusters.

Determining the *T* value is a simple task, but some trials still need to be undertaken in this regard.

This study proposes applying Manhattan L-function for flows to make the existing L-function adaptable to road networks, thereby yielding more practical results. Identifying aggregation in constrained flow space also help the setting of distribution sites, say the metro stations for humans and logistics centres for goods. Take the site selection of post office as an example, the aggregation centre or any other locations along AAA of express delivery flows are the alternative site location, and the aggregation scale indicates the distribution path length of most express delivery of the site. Future studies should focus on noise identification and consider the measurement of proximity between flows with different lengths and types for better detecting flow aggregation.

**Acknowledgments**

The authors thank the editor and the anonymous reviewers for their helpful comments on an earlier draft of this paper.

**Funding**

This work was supported by the National Natural Science Foundation of China (Grant Nos: 41525004 and 42071436).

**Disclosure Statement**

The authors declare no competing interests.

**Data availability statement**

The data and codes that support the findings of this study are available in 'figshare.com' with the identifier(s): https://doi.org/10.6084/m9.figshare.15086238.

**Figure Captions**

Figure 1 Flows in Manhattan flow space with maximum distance: (a) maximum distance between flows; (b) projection of a flow sphere in 4-D flow space

Figure 2 AAA of taxi OD flows in a road network. The red flows have a different movement type because the urban function of their origins and destinations differ from those of the dominant flows on AAA (colored black); the dominant flow in this cluster is from the recreational district to the commercial area; noise flows are from the business district to the food court.

Figure 3 Illustration of extracting AAA. In this case, $T = 2$, i.e., it picks top two core flows; the intersection of their *R-neighborhoods* will be the clustering range. Thus, the orange flows are in the key cluster; the blue flows are rejected. Highlight AAA by sketching the boundary of projection of key cluster on the network.

Figure 4 Extraction results with different $T$ values. The green points and orange points represent the O points and D points, respectively. $T = 1$ in (a); $T = 9$ in (b); $T = 21$ in (c); NAF, number of aggregated flows in the result

Figure 5 Simulated flow data; O points appear in green and D points are in orange.

Figure 6 Results with different clustering methods: (a) Manhattan flow DBSCAN; (b) Euclidean L-function; (c) Manhattan L-function. The O points appear in green and the D points in orange. The OD area within the dotted lines is the $\hat{R} - neighbors$ of the center flow.

Figure 7 Research area (Shunyi district) and the selected taxi OD flows

Figure 8 Results of the study case with Euclidean L-function: (a) L-function; (b) derivative of L-function

Figure 9 Results of the study case with Manhattan L-function: (a) L-function; (b) derivative of L-functionFigure 5

Figure 10 Clustering results of the maximal aggregation and secondary aggregation: (a) result with Manhattan flow DBSCAN; (b) result with ELF; (c) result with MLF ($T = 3$)

Figure 11 Secondary aggregation obtained using MLF and corresponding AAA; the flows in this cluster are mostly from the commercial area to the residential district.

Figure 12 Difference between the secondary aggregation obtained using ELF and MLF; the noise flows clustered with ELF were rejected with MLF. Some of these flows were from banks, schools, or bookstores and ended near clinics or business buildings, thereby confusing the main direction with this flow cluster.